\documentclass[9pt, twoside]{article}
\usepackage{csquotes}
 
\usepackage{graphics}
\usepackage{graphicx}
\usepackage{epsfig}
\usepackage{subfigure}

\usepackage{verbatim}
\usepackage{amsmath}
\usepackage{amssymb}
\usepackage{color}
\usepackage{bm}  
\usepackage{hyperref}
 \usepackage{stackrel}
 \usepackage{fullpage}

\def\##1{{\bf #1}}
\def\=#1{\underline{\underline #1}}
\def\~#1{{\tilde{\bf #1}}}

\def\r#1{(\ref{#1})}
\def\l#1{\label{#1}}

\def\eps{\varepsilon}
\def\epso{\eps_{\scriptscriptstyle 0}}

\def\muo{\mu_{\scriptscriptstyle 0}}
\def\ko{k_{\scriptscriptstyle 0}}
\def\etao{\eta_{\scriptscriptstyle 0}}

\def\inc{_{\rm inc}}
\def\sca{_{\rm sca}}
\def\exc{_{\rm exc}}
\def\ext{_{\rm ext}}

\def\smn{_{\rm smn}}
\def\mn{_{\rm mn}}

\def\epsr{\eps_{r}}
\def\mur{\mu_{r}}
\def\nr{\tilde{n}}

\def\les{\left[}
\def\ris{\right]}
\def\lec{\left\{}
\def\ric{\right\}}

\def\.{\mbox{ \tiny{$^\bullet$} }}

\def\ux{\hat{\#x}}
\def\uy{\hat{\#y}}

\def\ur{\hat{\#r}}
\def\uphi{\hat{\mbox{\boldmath$\phi$}}}
\def\utheta{\hat{\mbox{\boldmath$\theta$}}}

\def\pinm{\frac{m P_n^m(\cos\theta)}{\sin\theta}}
\def\taunm{\frac{dP_n^m(\cos\theta)}{d\theta}}
\def\one{^{(1)}}
\def\three{^{(3)}}

\def\cn{c_n}
\def\dn{d_n}
\def\en{e_n}
\def\pn{p_n}
\def\qn{q_n}
\def\sn{s_n}
\def\tn{t_n}
\def\Deltan{\Delta_n}

\def\fsc{\tilde{\alpha}}

\def\eomn{_{\stackrel[{\rm o}]{\rm e}{}{_{\rm mn}}}}

\begin{document}

\begin{center}

\textbf{Electromagnetic scattering by   homogeneous, isotropic, dielectric-magnetic sphere with topologically insulating surface states} \\

\textit{Akhlesh Lakhtakia}\\
{
The Pennsylvania State University, Department of Engineering Science and Mechanics, Nanoengineered Metamaterials Group,  University Park, PA 16802, USA}

\textit{Tom G. Mackay}\\
{
The Pennsylvania State University, Department of Engineering Science and Mechanics, Nanoengineered Metamaterials Group,  University Park, PA 16802, USA}\\
]{University of Edinburgh, School of Mathematics and
   Maxwell Institute for Mathematical Sciences,
Edinburgh EH9 3FD, Scotland, United Kingdom}\\

\end{center}

\begin{abstract} 
 The Lorenz--Mie formulation of electromagnetic scattering by a homogeneous, isotropic, dielectric-magnetic sphere was extended to incorporate topologically insulating surface states characterized by a surface admittance $\gamma$. Closed-form expressions were derived for the expansion coefficients of the scattered field phasors
in terms of those of the incident field phasors. These expansion coefficients were used to obtain analytical expressions for  the total scattering, extinction, forward scattering, and backscattering efficiencies  of the sphere. Resonances exist for relatively low values
of $\gamma$, when the sphere is either nondissipative or weakly dissipative. For large
 values of  $\gamma$, the scattering characteristics are close to that of a perfect electrically conducting sphere,
regardless of whether the sphere is composed of a dissipative or nondissipative material, and regardless of whether  that material supports planewave propagation with positive or negative phase velocity.

\end{abstract}

 \section{Introduction}\label{sec:intro}
Complete analytical treatment of electromagnetic planewave scattering  by a
homogeneous, isotropic, dielectric
sphere \cite{Logan}  can be traced back to an  {1890 paper of Lorenz \cite{Lorenz1890,Lorenz1898}},
though credit for that achievement is commonly given to a 1908 paper of Mie \cite{Mie1908}.
The incident, the scattered, and the internal fields are expanded as series of
 vector spherical wavefunctions \cite{Stratton1941,MF1953}, the surface of the sphere is taken to be charge free and current free, the standard boundary conditions of electromagnetics are imposed, and the orthogonality properties of the trigonometric functions and the associated Legendre functions are exploited. Extensions of this treatment to isotropic dielectric-magnetic spheres
\cite{Stratton1941,BH1983},  bi-isotropic spheres \cite{Bohren1974}, and some
 orthorhombic dielectric--magnetic spheres \cite{JL2014,JL2014c} have been reported.

Electrification of the surfaces of dielectric
 particles is commonly observed \cite{Masuda1976,Desch2002,Sow2011}. It too can
 be incorporated in the Lorenz--Mie formulation through a two-sided boundary
 condition involving a surface electric current density that is proportional to
 the tangential electric field \cite{BH1977}. This type of impedance boundary
 condition is the same as commonly used for carbon nanotubes \cite{Slepyan1999}
 and represents the formation of surface  states for electronic propagation \cite{Slepyan1999,Wei2004,Moradi2010}.

Surface states exist on topological insulators
as protected conducting states and are responsible for the characteristic electromagnetic responses of these
materials  \cite{Hasan2010}.  {Two classical-electromagnetic models have been proposed
for these materials as follows.
\begin{itemize}
\item[I.] The topological insulator
is an  achiral nonreciprocal bi-isotropic material characterized by three scalar 
constitutive parameters: the relative permittivity
$\epsr$, the relative permeability $\mur$, and the Tellegen nonreciprocity parameter $\gamma$ \cite{Qi}. The surface of a finite region occupied by the topological
insulator is charge-neutral and current-neutral.
\item[II.] The topological insulator is an
isotropic dielectric-magnetic material characterized by the relative permittivity
$\epsr$ and the relative permeability $\mur$, but its surface is endowed 
with surface charge and current densities quantitated through a non-null surface 
admittance $\gamma$ \cite{LM2015}. 
\end{itemize}
Model~I is physically inadequate because the essential {macroscopic} physics of topological insulation  occurs not inside a region but on the surface of that region. Indeed, the Tellegen
nonreciprocity  parameter
 disappears
from the Maxwell equations applicable to that region \cite{Pal}, and even leads by itself to
a contradiction \cite{IMA-JAM}. In contrast, the surface admittance of Model~II appears in the
boundary conditions, in consonance with the existence of \textit{surface} states.
 Let us  note, in passing, that any model of a topological insulator with $\gamma=0$ and a surface
conductivity derivable from a complex relative permittivity \cite{Autore} ignores the
existence of surface states, and therefore should be
valid only 
under the long-wavelength approximation.}

In this paper, we  {adopt Model~II to} extend the Lorenz--Mie formulation in order to encompass electromagnetic scattering by a {homogeneous, isotropic,} dielectric-magnetic sphere with topologically
insulating surface states. The incident field is not necessarily a plane wave, but its sources must lie outside the sphere
\cite{Wood,LI1983} and are assumed to be unaffected by the scattered field. The  {$\exp(-i\omega{t})$} time dependence is implicit, with $i=\sqrt{-1}$, $\omega$ as the angular frequency, and $t$ as time.  The free-space wavenumber $\ko = \omega \sqrt{\epso \muo}$ and  the free-space intrinsic impedance $\etao = \sqrt{\muo/\epso}$,
where $\epso$ and $\muo$ are the
permittivity and permeability of free space, respectively.
Vector quantities are displayed in bold typeface, with the superscript symbol $\hat{}$ denoting a unit vector.

\section{{Boundary-value problem}}\label{bvp}

Consider the sphere $r< a$ made of an isotropic dielectric-magnetic material with relative permittivity $\epsr$ and
relative permeability $\mur$. We also define a surface admittance $\gamma$
for use in boundary conditions at $r=a$,  {in accordance with the physically appropriate Model~II}. The external region $r>a$ is vacuous.  

\subsection{Incident electromagnetic field\label{incident}}

In a region that completely encloses the spherical surface $r=a$ but excludes the sources of the incident electromagnetic field, the
 incident electric and magnetic field phasors are represented  {as \cite{Wood,LI1983}}
\begin{eqnarray}
&&\nonumber
{\#E}\inc(\#r) = \sum_{s\in\left\{e,o\right\}}\sum^{\infty}_{n=1}\sum^n_{m=0}
\left\{D\mn\left[A\smn\one\,
{{\#M}\smn\one}(\ko\#r)
\right.\right.\\
&&\qquad\qquad\qquad\left.\left.+
B\smn\one\,
{{\#N}\smn\one}(\ko\#r)\right]\right\}\,,
\label{incE2}
\\&&
\nonumber
{\#B}\inc(\#r) = \frac{\ko}{i\omega}\,\sum_{s\in\left\{e,o\right\}}\sum^{\infty}_{n=1}\sum^n_{m=0}
\left\{D\mn\left[A\smn\one\,
{{\#N}\smn\one}(\ko\#r)
\right.\right.\\
&&\qquad\qquad\qquad\left.\left.+
B\smn\one\,
{{\#M}\smn\one}(\ko\#r)\right]\right\}\,,
\label{incB2}
\end{eqnarray}
where  the vector spherical  wavefunctions
${{\#M}\smn\one}(\ko\#r)$ and ${{\#N}\smn\one}(\ko\#r)$   
 {\cite{Stratton1941,MF1953}} are defined in the Appendix, and
the normalization factor
\begin{equation}
\displaystyle{
D\mn=(2-\delta_{\rm m0}){(2n+1)(n-m)!\over 4n(n+1)(n+m)!}
}
\end{equation}
employs the Kronecker delta $\delta_{\rm mm^\prime}$. The coefficients  ${{A}\smn\one}$ and ${{B}\smn\one}$ are presumed to be known.  {The functions 
${{\#M}\smn\one}(\ko\#r)$ are classified as toroidal and the functions
${{\#N}\smn\one}(\ko\#r)$ as poloidal \cite{Chan1961}.}

\subsection{Scattered electromagnetic field\label{scattered}}
The scattered electric and magnetic field phasors are represented  {as \cite{Waterman1965,Waterman1971}}
\begin{eqnarray}
&&\nonumber
{\#E}\sca(\#r) = \sum_{s\in\left\{e,o\right\}}\sum^{\infty}_{n=1}\sum^n_{m=0}
\left\{D\mn\left[A\smn\three\,
{{\#M}\smn\three}(\ko\#r)
\right.\right.\\
&&\qquad\qquad\qquad\left.\left.+
B\smn\three\,
{{\#N}\smn\three}(\ko\#r)\right]\right\}\,,\quad r>a\,,
\label{scaE2}
\\&&
\nonumber
{\#B}\sca(\#r) = \frac{\ko}{i\omega}\,\sum_{s\in\left\{e,o\right\}}\sum^{\infty}_{n=1}\sum^n_{m=0}
\left\{D\mn\left[A\smn\three\,
{{\#N}\smn\three}(\ko\#r)
\right.\right.\\
&&\qquad\qquad\qquad \left.\left.+
B\smn\three\,
{{\#M}\smn\three}(\ko\#r)\right]\right\}\,,\quad r>a\,,
\label{scaB2}
\end{eqnarray}
where  the vector spherical wavefunctions ${{\#M}\smn\three}(\ko\#r)$ and ${{\#N}\smn\three}(\ko\#r)$    {\cite{Stratton1941,MF1953}} are defined in the Appendix. The coefficients  ${{A}\smn\three}$ and ${{B}\smn\three}$ have to be determined.  {The functions 
${{\#M}\smn\three}(\ko\#r)$ are classified as toroidal and the functions
${{\#N}\smn\three}(\ko\#r)$ as poloidal.}

In the far zone,  the scattered electric field  may be approximated  {as \cite{Saxon1955}}
\begin{equation}
\#E\sca({\#r})\approx\#F\sca(\theta,\phi) \frac{\exp(i{\ko}r)}{r}\,
\end{equation}
and the scattered magnetic field as
\begin{equation}
\#B\sca({\#r})\approx\frac{\ko}{\omega}\ur\times\#F\sca(\theta,\phi) \frac{\exp(i{\ko}r)}{r}\,,
\end{equation}
where $\hat{\#r}=\#r/r$ and
\begin{eqnarray}
\nonumber
&&\#F\sca(\theta,\phi) = \ko^{-1}\,\sum_{s\in\left\{e,o\right\}}\sum^{\infty}_{n=1}\sum^n_{m=0}\left\{
(-i)^n\,D\mn\,\sqrt{n(n+1)}\,
\right.
\\
&&\qquad \left.
\left[-iA\smn\three\,
{{\bf C}\smn}(\theta,\phi)+
B\smn\three\,
{{\bf B}\smn}(\theta,\phi)\right]\right\}\,
\label{Fsca-def}
\end{eqnarray}
 is the vector far-field scattering amplitude. The angular harmonics
 ${{\bf B}\smn}(\theta,\phi)$ and ${{\bf C}\smn}(\theta,\phi)$  {\cite{MF1953}} are defined in the Appendix.

\subsection{Internal electromagnetic field\label{internal}}
The electric and magnetic field phasors excited inside the
{chosen sphere} are represented  {by \cite{Waterman1971}}
\begin{eqnarray}
&&\nonumber
{\# E}\exc({\#r}) =\sum_{s\in\left\{e,o\right\}}\sum^{\infty}_{n=1}\sum^n_{m=0}\left\{D\mn
 \left[\alpha\smn\,
{{\#M}\smn\one}(k{\#r})
\right.\right.\\
&&\qquad\qquad\qquad\left.\left.+
\beta\smn\,
{{\#N}\smn\one}(k{\#r})\right]\right\}\,, \quad r< a\,,
\label{Eint}
\\&&
\nonumber
{\#B}\exc({\#r}) = \frac{k}{i\omega}
 \sum_{s\in\left\{e,o\right\}}\sum^{\infty}_{n=1}\sum^n_{m=0}\left\{D\mn
 \left[\alpha\smn\,
{{\#N}\smn^{(1)}}(k{\#r})
\right.\right.
\\
&&\qquad\qquad\qquad
\left.\left.
+\beta\smn\,
{{\#M}\smn^{(1)}}(k{\#r})\right]\right\}\,, \quad r < a\,,
\label{Bint}
\end{eqnarray}
 the coefficients $\alpha\smn$ and $\beta\smn$ being unknown. The wavenumber $k=\nr\ko$,
 where $\nr =\sqrt{\epsr}\sqrt{\mur}$ is the refractive index.  {Equations~\r{Eint}
 and \r{Bint} are respectively similar to Eqs.~\r{incE2} and \r{incB2}, because the source-free region  $r <a$ contains the origin.}

\subsection{Boundary conditions \label{boundary}}

 {In accordance with Model~II,} the  boundary conditions appropriate for a topological insulator are as follows:
\begin{eqnarray}
\nonumber
&&\left.\begin{array}{l}
\ur\times\les\#E\inc(\#r)+\#E\sca(\#r)
-\#E\exc(\#r)\ris=\#0
\\[4pt]
\ur\times\lec\muo^{-1}\les\#B\inc(\#r)+\#B\sca(\#r)\ris-\left(\muo \mur\right)^{-1}
\#B\exc(\#r)\ric\\[4pt]
\qquad=-\gamma\ur\times\#E\exc(\#r)
\end{array}\right\}\,,
\\[5pt]
&&\qquad\qquad
{r=a}\,.
\label{bc}
\end{eqnarray}
 {The surface admittance $\gamma$ characterizing the topologically insulating
surface states} is very likely  dependent on the free-space wavenumber as
well as the constitution of the topological insulator,  {and it} may also
vary on the surface according to the local geometry. We hypothesize that a minimum radius of curvature is necessary
for the existence of a non-null admittance. 
For a homogeneous sphere, symmetry suggests that the surface admittance  does not vary on the surface.  Furthermore, we expect that $\gamma=0$ in the long-wavelength approximation ($\ko{a}\ll1$ and $\vert{k}\vert{a}\ll1$
\cite{vandeHulst})  because surface states will be either non-existent or inconsequential in the absence of a sufficient volume. This expectation is in accord with experimental results \cite{Autore}  on very thin films of the topologically insulating chalcogenide Bi$_2$Se$_3$,
those experimental results being evidently accounted for by a  surface conductivity arising from a permittivity, without requiring $\gamma\ne0$.

The boundary conditions \r{bc}  differ from their counterparts  \cite{BH1977}
\begin{eqnarray}
\nonumber
&&\left.\begin{array}{l}
\ur\times\les\#E\inc(\#r)+\#E\sca(\#r)
-\#E\exc(\#r)\ris=\#0
\\[4pt]
\ur\times\lec\muo^{-1}\les\#B\inc(\#r)+\#B\sca(\#r)\ris-\left(\muo \mur\right)^{-1}
\#B\exc(\#r)\ric\\[4pt]
\qquad=\tilde{\sigma} \left(\=I-\ur\ur\right)
\.\#E\exc(\#r)
\end{array}\right\}\,,
\\[5pt]
&&\qquad\qquad
{r=a}\,,
\label{bc-chargedsph}
\end{eqnarray}
that
prevail on the surface of a charged sphere, where $\=I$ is the identity dyadic \cite{Chen1983} and the frequency-dependent
 surface conductivity
$\tilde{\sigma}$    incorporates the charging of the surface $r=a$.  {The difference can be appreciated by noting that $\ur\times\#A\ne \left(\=I-\ur\ur\right)\.\#A$ for an arbitrary vector
$\#A=A_r\ur+A_\theta\utheta+A_\phi\uphi$; indeed,  $\ur\times\#A=-\utheta A_\phi +\uphi A_\theta$ but $\left(\=I-\ur\ur\right)\.\#A=\utheta A_\theta+\uphi A_\phi$ on the surface of the sphere. The use of Eqs.~\r{bc-chargedsph} in lieu of Eqs.~\r{bc} is inadmissible for a topological insulator, as is clear from Sec.~2\ref{comp-chargedsph}.}

The twin boundary conditions \r{bc}  also differ from the sole boundary condition
\cite{Garbacz1964,Medgyesi1985}
\begin{eqnarray}
\nonumber
&&\tilde{\sigma}\ur\times\les\#E\inc(\#r)+\#E\sca(\#r)\ris=\\[4pt]
&&\qquad-\muo^{-1}\left(\=I-\ur\ur\right)
\.
\les\#B\inc(\#r)+\#B\sca(\#r)\ris\,,\quad
{r=a}\,,
\label{bc-impsph}
\end{eqnarray}
that is taken to prevail on the surface of an impedance sphere.
 {On taking the cross product of both sides with $\ur$,
the boundary condition \r{bc-impsph} is equivalently written as 
\begin{eqnarray}
\nonumber
&&\tilde{\sigma}\left(\=I-\ur\ur\right)\.\les\#E\inc(\#r)+\#E\sca(\#r)\ris=\\[4pt]
&&\qquad\muo^{-1}\ur\times
\les\#B\inc(\#r)+\#B\sca(\#r)\ris\,,\quad
{r=a}\,.
\label{bc-impsph-equiv}
\end{eqnarray}
The use of an impedance boundary condition results in loss of information about the internal fields
${\# E}\exc({\#r})$ and ${\# B}\exc({\#r})$, and is generally adopted for perfectly conducting objects with somewhat rough \cite{Hiatt1960}  or coated surfaces \cite{Weston1962}.
An impedance
boundary condition cannot represent the  surface states of a topological insulator as becomes clear in Sec.~2\ref{comp-impsph}.}

\subsection{Solution of boundary-value problem\label{solution}}

After substituting Eqs.~\r{incE2}, \r{scaE2}, and \r{Eint} in Eq.~\r{bc}$_1$, and after exploiting the orthogonality properties
of the trigonometric functions over $0\leq\phi\leq2\pi$ and of the associated Legendre functions over $0\leq\theta\leq\pi$,
we obtain the  simple algebraic equations
\begin{equation}
\left.\begin{array}{l}
A\smn\one\, j_n(\xi) + A\smn\three\, h\one_n(\xi)
=\alpha\smn\, j_n(\nr\xi)
\\[5pt]
\nr\les
B\smn\one\, \psi_n\one(\xi) + B\smn\three \,\psi_n\three(\xi)\ris
=\beta\smn\, \psi_n\one(\nr\xi)
\end{array}\right\}\,,
\label{eqs1&2}
\end{equation}
where $\xi=\ko{a}$.
Likewise, after substituting Eqs.~\r{incB2}, \r{scaB2}, \r{Eint}, and \r{Bint} in Eq.~\r{bc}$_2$, and after exploiting the orthogonality properties
of the trigonometric functions over $0\leq\phi\leq2\pi$ and of the associated Legendre functions over $0\leq\theta\leq\pi$,
we obtain the following simple algebraic equations:
\begin{equation}
\left.\begin{array}{l}
\mur\les
A\smn\one\, \psi_n\one(\xi) + A\smn\three\, \psi_n\three(\xi)\ris\\[5pt]
\qquad
=\les\alpha\smn-i\left(\etao\gamma\nr/\epsr\right)\beta\smn\ris
\psi_n\one(\nr\xi)
\\[5pt]
\frac{\nr}{\epsr}\les
B\smn\one\, j_n(\xi) + B\smn\three\, h\one_n(\xi)\ris\\[5pt]
\qquad
=\les\beta\smn-i\left(\etao\gamma\nr/\epsr\right)\alpha\smn\ris
j_n(\nr\xi)
\end{array}\right\}\,.
\label{eqs3&4}
\end{equation}

Equations~\r{eqs1&2} and \r{eqs3&4} are straightforward to solve for $A\smn\three$, $B\smn\three$,
$\alpha\smn$, and $\beta\smn$ in terms of $A\smn\one$ and $B\smn\one$. As our interest lies in the
scattered fields, we are content to state that
\begin{equation}
\left.\begin{array}{l}
\label{eq17}
A\smn\three = \cn \, A\smn\one + \dn \, B\smn\one\\[5pt]
B\smn\three=-\dn\, A\smn\one+\en\,B\smn\one
\end{array}\right\}\,,
\end{equation}
where
\begin{eqnarray}
\nonumber
&&
\Deltan\cn = -\sn\qn
\\[5pt]
&&\qquad
+\left(\etao\gamma\right)^2\mur\,
j_n(\xi)\psi_n\three(\xi)\,
j_n(\nr\xi)\psi_n\one(\nr\xi)\,,
\label{cn-def}
\\[5pt]
&&
\Deltan\dn= \left(\etao\gamma \mur/\xi\right)\,
j_n(\nr\xi)\psi_n\one(\nr\xi)\,,
\label{dn-def}
\\[5pt]
\nonumber
&&
\Deltan\en = -\pn\tn
\\[5pt]
&&\qquad
+\left(\etao\gamma\right)^2\mur\,
h_n\one(\xi)\psi_n\one(\xi)\,
j_n(\nr\xi)\psi_n\one(\nr\xi)\,,
\label{en-def}
\\[5pt]
\nonumber
&&
\Deltan  = \qn\tn
\\[5pt]
&&\qquad
-\left(\etao\gamma\right)^2\mur\,
h_n\one(\xi)\psi_n\three(\xi)\,
j_n(\nr\xi)\psi_n\one(\nr\xi)\,,
\label{Deltan-def}
\\[5pt]
&&
\pn=\epsr\,j_n(\nr\xi)\psi_n\one(\xi)
-j_n(\xi)\psi_n\one(\nr\xi)\,,
\label{pn-def}
\\[5pt]
&&
\qn=\epsr\,j_n(\nr\xi)\psi_n\three(\xi)
-h_n\one(\xi)\psi_n\one(\nr\xi)\,,
\label{qn-def}
\\[5pt]
&&
\sn=\mur\,j_n(\nr\xi)\psi_n\one(\xi)
-j_n(\xi)\psi_n\one(\nr\xi)\,,
\label{sn-def}
\\[5pt]
&&
\tn=\mur\,j_n(\nr\xi)\psi_n\three(\xi)
-h_n\one(\xi)\psi_n\one(\nr\xi)\,.
\label{tn-def}
\end{eqnarray}
 {Toroidal-poloidal mixing on scattering is signified by $d_n\ne0$.}

 {\subsection{Comparison with Model~I}\label{comp-Model-I}
Planewave scattering by a homogeneous, isotropic, dielectric-magnetic sphere with topologically
insulating surface states was solved
by Ge \textit{et al.}  \cite{Ge2015} recently using Model~I. However,
that boundary-value problem is a simple specialization 
of a more general one for a bi-isotropic sphere,  solved four decades earlier
by Bohren  \cite{Bohren1974}. The specialization requires
setting $\alpha+\beta=0$ in Ref.~\cite{Bohren1974}. }

 {Anyhow, we have verified that application of Model~I also
yields Eqs.~\r{eq17}--\r{tn-def}. 
That both Models~I and II yield the same electromagnetic field scattered by a finite region occupied by a topological insulator has been also noted for the planewave reflection and
refraction due to a half space occupied by a homogeneous, isotropic, dielectric-magnetic material
with topologically
insulating surface states 
\cite{LM2016}. Thus,
the two models lead to different boundary-value problems for scattering
by a finite region occupied by a topological insulator, {but} the scattered field remains the same.
This conclusion, however, does not affect the physical deficiency inherent in Model~I, because  surface states reside on a surface (and can therefore impact the
boundary conditions through a surface admittance) whereas the Tellegen nonreciprocity
parameter is a constitutive parameter that must hold in a region bounded by that surface.
The experimental results of Autore \textit{et al.} \cite{Autore} clearly show that  surface states  vanish
when the volume is very tiny, but a constitutive parameter has to be valid regardless of the volume.}

 {\subsection{Comparison with charged sphere}\label{comp-chargedsph}
For a charged sphere \cite{BH1977}, Eqs.~\r{bc-chargedsph} have to be used in lieu
of Eqs.~\r{bc} but the remainder of the analytical treatment remains the same.
As a result, for use in Eq.~\r{eq17} the following expressions are obtained:
\begin{eqnarray}
&&
\cn = - \frac
{i\etao\tilde{\sigma}\mur\xi j_n(\xi) j_n(\nr\xi) + s_n}
{i\etao\tilde{\sigma}\mur\xi h_n\one(\xi) j_n(\nr\xi) + t_n}\,,
\\
&&
\dn=0\,,
\\
&&
\en = -\frac
{i\etao\tilde{\sigma}\psi_n\one(\xi)\psi_n\one(\nr\xi)+\xi p_n}
{i\etao\tilde{\sigma}\psi_n\three(\xi)\psi_n\one(\nr\xi)+\xi q_n}\,.
\end{eqnarray}
Clearly,  toroidal-poloidal mixing does not occur
on scattering by a charged sphere---unlike for an isotropic dielectric-magnetic sphere with topologically
insulating surface states.
Hence, a surface-conductivity model \cite{Autore} is inadmissible
for a dielectric-magnetic sphere with topologically insulating surface states.}

 {\subsection{Comparison with impedance sphere}\label{comp-impsph}
For an impedance sphere \cite{Hiatt1960,Weston1962}, either Eq.~\r{bc-impsph} 
or Eq.~\r{bc-impsph-equiv} has to be used in lieu
of Eqs.~\r{bc}, and the analytical treatment is similar.
As a result, for use in Eq.~\r{eq17} the following expressions are obtained:
\begin{eqnarray}
\label{eq29}
&&
\cn = - \frac
{i\etao\tilde{\sigma}\xi j_n(\xi)  + \psi_n\one(\xi)}
{i\etao\tilde{\sigma}\xi h_n\one(\xi)  + \psi_n\three(\xi)}\,,
\\
&&
\dn=0\,,
\\
&&
\label{eq31}
\en = -\frac
{i\etao\tilde{\sigma}\psi_n\one(\xi) -\xi j_n(\xi)}
{i\etao\tilde{\sigma}\psi_n\three(\xi) -\xi h_n\one(\xi)}\,.
\end{eqnarray}
Thus,  as toroidal-poloidal mixing does not occur
on scattering by an impedance sphere, an impedance
boundary condition is inadmissible
for a dielectric-magnetic sphere with topologically insulating surface states.}

 {\subsection{Comparison with perfect electrically conducting  sphere}\label{comp-PECsph}
For a perfect electrically conducting (PEC) sphere \cite{BSU}, the limit $\tilde{\sigma}\to \infty$
must be taken in Eqs.~\r{eq29}--\r{eq31} to obtain
\begin{eqnarray}
\label{eq32}
&&
\cn = - \frac
{j_n(\xi)   }
{h_n\one(\xi)   }\,,
\\
&&
\dn=0\,,
\\
&&
\label{eq34}
\en = -\frac
{\psi_n\one(\xi)  }
{\psi_n\three(\xi)  }\,.
\end{eqnarray}
}

\subsection{Planewave scattering}\label{pws}
Suppose that the incident electromagnetic field is a plane wave. Without loss of generality,
we can take it to be traveling along the $+z$ axis; hence,
\begin{equation}
\left.\begin{array}{l}
\#E\inc(\#r)= \ux\exp\left(i\ko{z}\right)\\[5pt]
\#B\inc(\#r)= \frac{\ko}{\omega}\uy\exp\left(i\ko{z}\right)
\end{array}\right\}\,,
\end{equation}
so  {that \cite{Stratton1941,MF1953,Waterman1965}}
\begin{equation}
\left.\begin{array}{l}
A\smn\one=2n(n+1)i^n \, \delta_{so}\,\delta_{m1}
\\[5pt]
B\smn\one=2n(n+1)i^{n-1} \,\delta_{se}\,\delta_{m1}
\end{array}\right\}\,.
\label{plw-coeffs}
\end{equation}

Accordingly, the non-zero coefficients in the expansions of the scattered electric and magnetic fields are as  follows:
\begin{equation}
\left.\begin{array}{l}
A_{o1n}\three=\cn\,A_{o1n}\one
\\[5pt]
A_{e1n}\three=\dn\,B_{e1n}\one
\\[5pt]
B_{o1n}\three=-\dn\,A_{o1n}\one
\\[5pt]
B_{e1n}\three=\en\,B_{e1n}\one
\end{array}\right\}\,.
\label{plw-sol}
\end{equation}
Using these coefficients in Eq.~\r{Fsca-def}, we can obtain the differential scattering efficiency  {\cite{Saxon1955}}
\begin{equation}\label{sigmad-def}
Q_{\rm D}(\theta,\phi)\triangleq\frac{4}{a^2} {\#F\sca(\theta,\phi)\.\#F\sca^\ast(\theta,\phi)} \,
\end{equation}
along any radial direction specified by the angles $\theta$ and $\phi$. The extinction efficiency   {\cite{deHoop1958,Saxon1955}}
\begin{eqnarray}
\label{Qext-def0}
Q\ext& =&\frac{4}{\ko{a^2}} {\rm Im} \lec{\#F\sca(0,0)\.\ux}\ric
\\[5pt]
&=&-\frac{2}{\xi^2}{\rm Re}\lec
\sum_{n=1}^{\infty} \les (2n+1)\left(\cn+\en\right)\ris\ric\,,
\label{Qext-def}
\end{eqnarray}
the forward-scattering  efficiency  {\cite{BSU,Saxon1955}}
\begin{eqnarray}
\label{Qf-def0}
Q_{\rm f}&\triangleq&Q_{\rm D}(0,0)
\\[5pt]
&=&\frac{1}{\xi^2}\Big\vert
\sum_{n=1}^{\infty} \les (2n+1)\left(\cn+\en\right)\ris\Big\vert^2\,,
\label{Qf-def}
\end{eqnarray}
the backscattering efficiency  {\cite{BSU}}
\begin{eqnarray}
\label{Qb-def0}
Q_{\rm b}&\triangleq&Q_{\rm D}(\pi,\pi)
\\[5pt]
\nonumber
&=&\frac{1}{\xi^2}\left\{\Big\vert
\sum_{n=1}^{\infty} \les (-1)^n(2n+1)\left(\cn-\en\right)\ris\Big\vert^2
\right.
\\[5pt]
&&\left.\qquad
+ 4\Big\vert
\sum_{n=1}^{\infty} \les (-1)^n(2n+1)\dn\ris\Big\vert^2\right\}
\,,
\label{Qb-def}
\end{eqnarray}
and the total scattering efficiency  {\cite{Waterman1971,Saxon1955}}
\begin{eqnarray} 
\label{tse-def0}
Q\sca&\triangleq&\frac{1}{4\pi}\,\int_{\phi=0}^{2\pi}\int_{\theta=0}^{\pi}\,Q_{\rm D}(\theta,\phi)\sin\theta\,d\theta\,d\phi\,
\\[5pt]
&=&\frac{2}{\xi^2}
\sum_{n=1}^{\infty}
\lec (2n + 1)\les \vert\cn\vert^2+2\vert\dn\vert^2+\vert\en\vert^2 \ris\ric
\,
\l{tse-def}
\end{eqnarray}
can be calculated.  {Let us note that the presence of $\sin\theta$
in the integrand on the right side of Eq.~\r{tse-def0} ensures that the
magnitudes of $Q_{\rm f}$ and $Q_{\rm b}$ do not affect the magnitude
of $Q\sca$.}

\section{Results and Discussion}
The effect of the topologically insulating surface states is seen clearly by setting $\epsr=\mur=1$; then,
$\cn\ne0$, $\dn\ne0$, and $\en\ne0$
 so long as $\gamma\ne0$. Thus, these surface states by themselves cause
 scattering,
 
 {Intrinsic topological insulators are characterized by $\gamma=\pm \fsc/\etao$  \cite{Qi},
where $\fsc =\left(q_e^2/\hbar{c}\right)/4\pi\epso$ is the (dimensionless) fine structure
constant,  $q_e=1.6\times10^{-19}$~C is the quantum of charge, $\hbar$ is the reduced Planck constant, and 
$c=1/\sqrt{\epso\muo}$ is the speed of light in free space. A very thin coating of a magnetic material is often used to realize
$\gamma=(2q+1)\fsc/\etao$, $q\in\lec0,\pm1,\pm2,\pm3,\dots\ric$ \cite{Qi}. 
Thus, both negative
and positive values of $\gamma$ are possible. The} replacement of $\gamma$ by $-\gamma$ does not affect $\cn$ and $\en$,
but definitely alters the sign of $\dn$, as is clear from Eqs.~\r{cn-def}--\r{Deltan-def}.
Concurrently, this replacement does not affect
$A_{o1n}\three$ and $B_{e1n}\three$, but it does alter the signs of
$A_{e1n}\three$ and $B_{o1n}\three$,
according to Eqs.~\r{plw-sol}.  In other words, the effect of
$\gamma$  on the depolarized scattered fields in planewave
scattering is significant. Nevertheless, $Q\ext$ and
$Q_{\rm f}$ are not directly affected by $\dn$, whereas  $Q\sca$ and
$Q_{\rm b}$ are not affected by the sign of $\dn$.
Therefore, replacement of $\gamma$ by $-\gamma$ does not
affect any of these four efficiencies. In the remainder of this section, we have confined
ourselves to non-negative $\gamma$.

When $\gamma=0$, i.e., in the absence of topologically insulating surface states,
Eqs.~\r{cn-def}--\r{Deltan-def} yield
\begin{eqnarray}
&&
\cn = -\sn/\tn
\,,
\label{cn0-def}
\\[5pt]
&&
\dn=0\,,
\label{dn0-def}
\\[5pt]
&&
\en = -\pn/\qn
\,.
\label{en0-def}
\end{eqnarray}
Indeed, $-\cn$ and $-\en$ are then, respectively, equal to the coefficients $b_n$ and $a_n$ of Bohren \& Huffman \cite[Eqs.~(4.53)]{BH1983} for scattering by
isotropic dielectric-magnetic spheres.

As $ \gamma $ increases beyond a sufficiently high value, Eqs.~\r{cn-def}--\r{Deltan-def} yield
\begin{eqnarray}
&&
\cn \rightarrow-
j_n(\xi)/
h_n\one(\xi)
\,,
\label{cninf-def}
\\[5pt]
&&
\dn\rightarrow -1/
\etao\gamma\xi  \,
h_n\one(\xi)\psi_n\three(\xi)\rightarrow 0
\,,
\label{dninf-def}
\\[5pt]
&&
\en \rightarrow-
\psi_n\one(\xi)/
\psi_n\three(\xi)
\,.
\label{eninf-def}
\end{eqnarray}
 {Thus, at very high values of $\gamma$,  the right sides of Eqs.~\r{cn-def}--\r{Deltan-def} tend towards those of Eqs.~\r{eq32}--\r{eq34}, the toroidal-poloidal mixing tends to vanish,
and the isotropic dielectric-magnetic sphere with topologically
insulating surface states tends to scatter like a PEC sphere.}

 {We carried out a parametric study to numerically assess the effect of $\gamma$ on scattering. We chose $\vert\epsr\vert\approx 3$, which is reasonable for many materials
in the optical regime. We also chose $\vert\mur\vert\approx 1.3$, which is quite in keeping
with ongoing efforts in the area of optical magnetism \cite{optmag1,optmag2}. Finally, we set
$\vert\gamma\vert \leq 1000 \fsc/\etao$, a very wide span for currently researched chalcogenide topological insulators but not inconceivable as the presently infant field of topological insulators grows to encompass mixed materials and new material compositions.}

\begin{figure}[!ht]
\centering
\includegraphics[width=14cm]{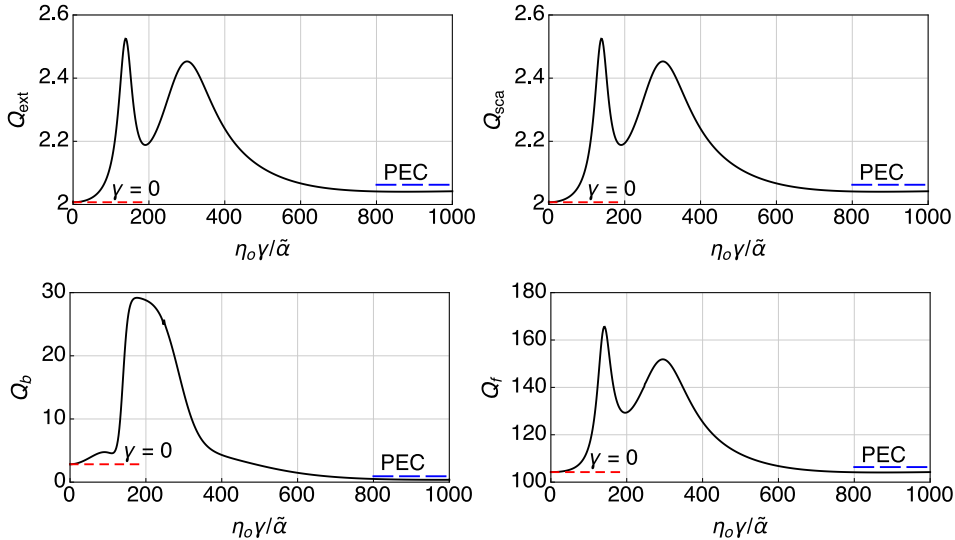}
 \caption{\label{Fig1} $Q\ext$, $Q\sca$, $Q_{\rm b}$, and $Q_{\rm f}$ as functions of
  $\etao\gamma/\fsc$   for a sphere ($\xi=10$, $\epsr=3$, $\mur=1.3$) embedded
  in free space. Values for $\gamma=0$ and a PEC sphere are
  also indicated.
  }
\end{figure}

 {The significance of the limits $\gamma\to0$ and $\gamma\to\infty$ is} evident in Fig.~\ref{Fig1}, wherein
$Q\ext$, $Q\sca$, $Q_{\rm b}$, and $Q_{\rm f}$ are plotted as functions of
  $\etao\gamma/\fsc$ for a nondissipative sphere ($\epsr=3$, $\mur=1.3$) of size parameter
  $\xi=10$. Two resonances are evident in these plots for  {$\etao\gamma/\fsc<400$}. As $\gamma$ increases further,  all four efficiencies approach their counterparts for
  a PEC sphere \cite{BSU}. Thus,
  a sphere with topologically insulating surface states becomes
  perfect electrically conducting  in the limit $\gamma\to\infty$, with Fig.~\ref{Fig1} indicating
  that this transition effectively happens for   {$\etao\gamma/\fsc\approx1000$}.

\begin{figure}[!ht]
\centering
\includegraphics[width=14cm]{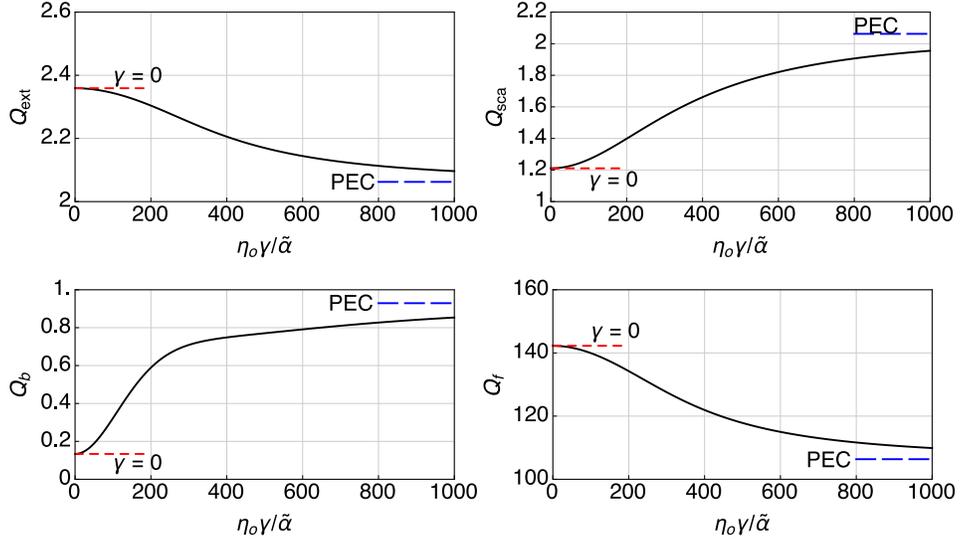}
 \caption{\label{Fig2} Same as Fig.~\ref{Fig1}, except that
 $\epsr=3+i0.1$ and $\mur=1.3+i0.05$.
  }
\end{figure}

  When the sphere material in Fig.~\ref{Fig1} is made dissipative by the addition of positive imaginary parts to $\epsr$ and $\mur$, the resonances broaden and eventually disappear. However,
  as shown in Fig.~\ref{Fig2} for $\epsr=3+i0.1$ and $\mur=1.3+i0.05$
  the transition to a PEC sphere still occurs as $\gamma$ increases.

\begin{figure}[!ht]
\centering
\includegraphics[width=14cm]{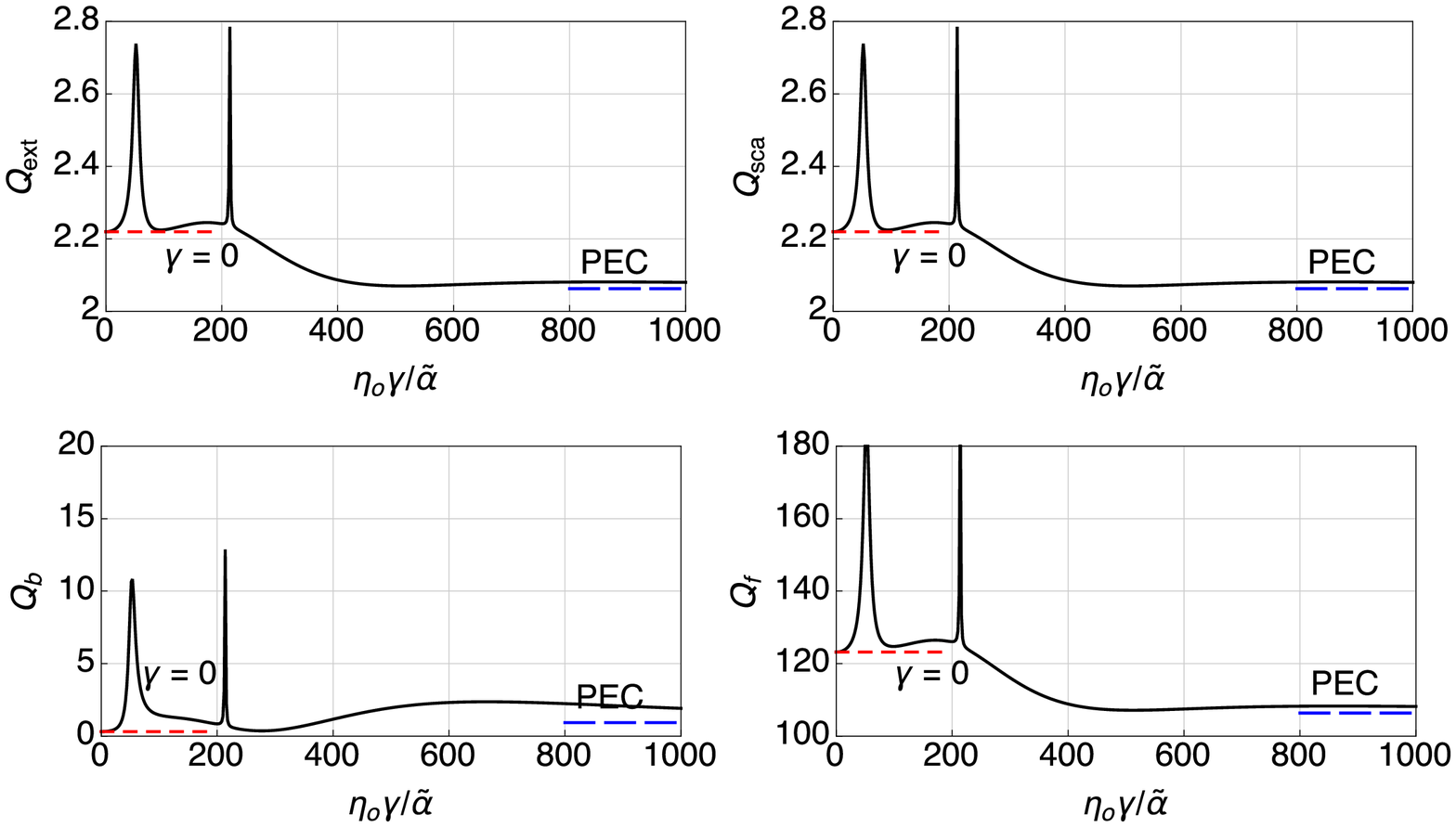}
 \caption{\label{Fig3}  Same as Fig.~\ref{Fig1}, except that
 $\epsr=-3$ and $\mur=-1.3$.
  }
\end{figure}

\begin{figure}[!ht]
\centering
\includegraphics[width=14cm]{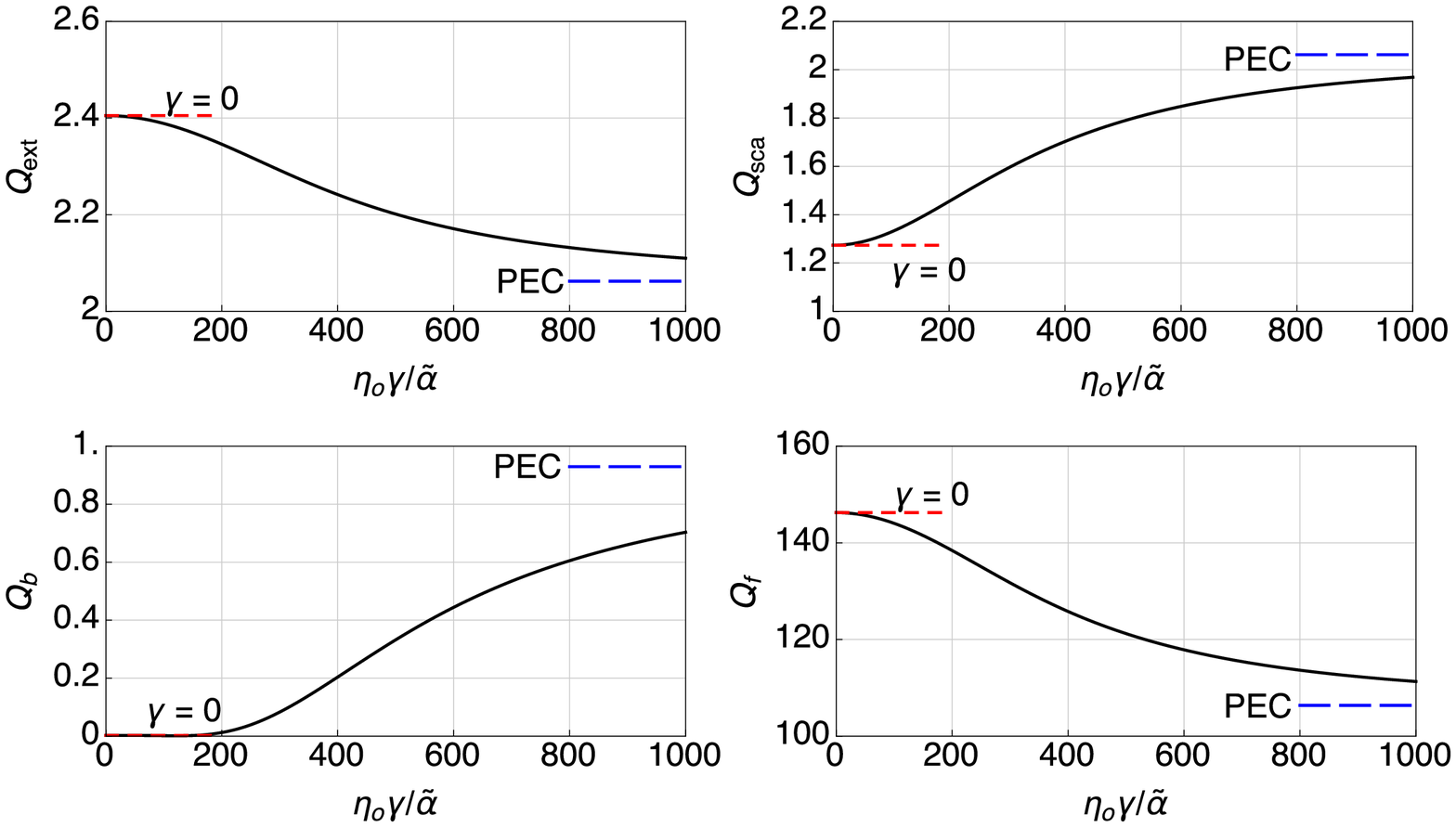}
 \caption{\label{Fig4}  Same as Fig.~\ref{Fig3}, except that
 $\epsr=-3+i0.1$ and $\mur=-1.3+i0.05$.
  }
\end{figure}

The data in Figs.~\ref{Fig1} and \ref{Fig2} were calculated for spheres
of materials that allow planewave propagation with positive phase velocity (PPV);
i.e., the phase velocity of a plane wave is co-parallel with the time-averaged
Poynting vector \cite{LMWaeu}. However, if the signs of the real parts of
both $\epsr$ and $\mur$ were to be negative, the phase velocity of a
plane wave will be anti-parallel with the time-averaged Poynting vector,
so that planewave propagation would occur with negative phase velocity (NPV).
Figures~\ref{Fig3} and \ref{Fig4} present data for NPV spheres with
topologically insulating surface states. Compared to Fig.~\ref{Fig1} drawn 
for  $\epsr=3$ and $\mur=1.3$, the resonances in  Fig.~\ref{Fig3} drawn 
for  $\epsr=-3$ and $\mur=-1.3$ are much sharper. However, the incorporation of
absorption makes the resonances disappear in Fig.~\ref{Fig4} for
 $\epsr=-3+i0.1$ and $\mur=-1.3+i0.05$. As $\gamma$ increases further,  all four efficiencies of an NPV sphere with  topologically insulating surface states approach their counterparts for a PEC sphere, just as for a PPV sphere with   topologically insulating surface states.

 {Rayleigh scattering by a  homogeneous, isotropic, dielectric-magnetic sphere with  topologically insulating surface states requires a comment. Formal expressions for the coefficients
$c_1$, $d_1$, and $e_1$ can be obtained in the limit $\xi\to0$ from Eqs.~\r{cn-def}--\r{tn-def}. But the experimental results of Autore \textit{et al.} \cite{Autore} clearly show that  surface states  vanish
when the volume is very tiny, indicating that $\gamma=0$
 in the long-wavelength approximation.}

\section{Concluding Remarks}\label{conc}

In the foregoing analysis, the incident, scattered, and internal field phasors were expanded
in terms of vector spherical wavefunctions, for  
electromagnetic scattering by a homogeneous, isotropic, dielectric-magnetic sphere with  topologically insulating surface states characterized by a surface admittance $\gamma$. 
Closed-form expressions were derived for the expansion coefficients of the scattered field phasors
in terms of those of the incident field phasors.

Numerical studies demonstrated the presence of resonances due to relatively low values
of $\gamma$, when the sphere is composed of a nondissipative or weakly dissipative material. Furthermore,
 the total scattering, extinction, forward scattering, and backscattering efficiencies  of the sphere {become} indistinguishable from those of a perfect electrically conducting sphere for sufficiently large values of  $\gamma$, regardless of whether the sphere is composed of a dissipative or nondissipative material, and regardless of whether  that material supports planewave propagation with positive or negative phase velocity.

\section*{Appendix}
The vector spherical wavefunctions regular at the origin are defined as \cite{Stratton1941,MF1953}
\begin{eqnarray}
\nonumber
&&{\#M}\eomn\one(\ko{\#r}) =
\mp\utheta \pinm j_n(\ko r)
{\lec\begin{array}{c}{\sin(m\phi)}\\{\cos(m\phi)}\end{array}\ric}
\\[5pt]
&&\qquad-\uphi \taunm j_n(\ko r)
{\lec\begin{array}{c}{\cos(m\phi)}\\{\sin(m\phi)}\end{array}\ric}\,
\end{eqnarray}
and
\begin{eqnarray}
\nonumber
&&
{\#N}\eomn\one(\ko{\#r}) =
\ur \,n(n+1) P_n^m(\cos\theta)\frac{j_n(\ko r)}{\ko r}
{\lec\begin{array}{c}{\cos(m\phi)}\\{\sin(m\phi)}\end{array}\ric}
\\[5pt]
\nonumber
&&\qquad+\utheta\taunm\frac{\psi_n\one(\ko r)}{\ko r}
{\lec\begin{array}{c}{\cos(m\phi)}\\{\sin(m\phi)}\end{array}\ric}
\\[5pt]
&&\qquad\mp  \uphi \pinm\frac{\psi_n\one(\ko r)}{\ko r}
{\lec\begin{array}{c}{\sin(m\phi)}\\{\cos(m\phi)}\end{array}\ric}\,,
\end{eqnarray}
whereas the ones regular at infinity are defined as
\begin{eqnarray}
\nonumber
&&
{\#M}\eomn\three(\ko{\#r}) =
\mp\utheta \pinm h_n\one(\ko r)
{\lec\begin{array}{c}{\sin(m\phi)}\\{\cos(m\phi)}\end{array}\ric}
\\[5pt]
&&\qquad-\uphi \taunm h_n\one(\ko r)
{\lec\begin{array}{c}{\cos(m\phi)}\\{\sin(m\phi)}\end{array}\ric}\,
\end{eqnarray}
and
\begin{eqnarray}
\nonumber
&&
{\#N}\eomn\three(\ko{\#r}) =
 \ur \,n(n+1) P_n^m(\cos\theta)\frac{h_n\one(\ko r)}{\ko r}
{\lec\begin{array}{c}{\cos(m\phi)}\\{\sin(m\phi)}\end{array}\ric}
\\[5pt]
\nonumber
&&\qquad+\utheta\taunm\frac{\psi_n\three(\ko r)}{\ko r}
{\lec\begin{array}{c}{\cos(m\phi)}\\{\sin(m\phi)}\end{array}\ric}
\\[5pt]
&&\qquad\mp   \uphi \pinm\frac{\psi_n\three(\ko r)}{\ko r}
{\lec\begin{array}{c}{\sin(m\phi)}\\{\cos(m\phi)}\end{array}\ric}\,.
\end{eqnarray}
In these expressions,
\begin{equation}
\left.\begin{array}{l}
\psi_n\one(w)= \frac{d}{dw}\left[w\,j_n(w)\right]
\\[5pt]
\psi_n\three(w)= \frac{d}{dw}\left[w\,h_n\one(w)\right]
\end{array}\right\}\,,
\end{equation}
$j_n(\.)$ denotes the spherical Bessel function of order $n$,
$h_n\one(\.)$ denotes the spherical Hankel function of the first kind and order $n$,
and $P_n^m(\.)$ is the associated
Legendre function of order $n$ and degree $m$.

The angular harmonics used in Eq.~\r{Fsca-def} are defined as \cite{MF1953}
\begin{eqnarray}
\nonumber
{\#B}\eomn
(\theta,\phi) &=& \frac{1}{\sqrt{n(n+1)}}\left[
\utheta \taunm
{\lec\begin{array}{c}{\cos(m\phi)}\\{\sin(m\phi)}\end{array}\ric}\right.
\\
&&\left.\mp\uphi \pinm
{\lec\begin{array}{c}{\sin(m\phi)}\\{\cos(m\phi)}\end{array}\ric}\right]
\end{eqnarray}
and
\begin{eqnarray}
\nonumber
{\#C}\eomn
(\theta,\phi) &=& \frac{1}{\sqrt{n(n+1)}}\left[
\mp\utheta \pinm
{\lec\begin{array}{c}{\sin(m\phi)}\\{\cos(m\phi)}\end{array}\ric}\right.
\\
&&\left.-\uphi \taunm
{\lec\begin{array}{c}{\cos(m\phi)}\\{\sin(m\phi)}\end{array}\ric}\right]\,.
\end{eqnarray}

The following identities are useful for various derivations:
\begin{equation}
\frac{P_n^1(\cos\theta)}{\sin\theta}\Big\vert_{\theta=0} = \frac{dP_n^1(\cos\theta)}{d\theta}\Big\vert_{\theta=0}= n(n+1)/2\,,
\end{equation}
\begin{equation}
\frac{P_n^1(\cos\theta)}{\sin\theta}\Big\vert_{\theta=\pi} =- \frac{dP_n^1(\cos\theta)}{d\theta}\Big\vert_{\theta=\pi}=
(-1)^{n+1} n(n+1)/2\,,
\end{equation}
\begin{eqnarray}
\nonumber
&&
\int_{0}^{\pi} \les \frac{mP_n^m(\cos\theta)}{\sin\theta} \frac{dP_{n^\prime}^m(\cos\theta)}{d\theta}
+ \frac{mP_{n^\prime}^m(\cos\theta)}{\sin\theta} \frac{dP_{n}^m(\cos\theta)}{d\theta}\ris
\\[4pt]
&&\qquad\quad\times\sin\theta\,d\theta
=0\,,
\end{eqnarray}
and
\begin{eqnarray}
\nonumber
&&
\int_{0}^{\pi} \les
\frac{mP_n^m(\cos\theta)}{\sin\theta} \frac{mP_{n^\prime}^m(\cos\theta)}{\sin\theta} +
\frac{dP_{n}^m(\cos\theta)}{d\theta} \frac{dP_{n^\prime}^m(\cos\theta)}{d\theta}\ris
\\[4pt]
&&\qquad\quad\times\sin\theta\,d\theta=
\frac{2}{2n+1}\frac{(n+m)!}{(n-m)!}\frac{(n+1)!}{(n-1)!}\,\delta_{nn^\prime}\,.
\end{eqnarray}

 \vspace{0.5cm}

\noindent{\it Acknowledgment.}   AL thanks the Charles Godfrey Binder Endowment
at Penn State for ongoing support of his research activities. {TGM acknowledges the support of EPSRC grant EP/M018075/1.}

\end{document}